\documentclass[aps,prb,twocolumn,superscriptaddress,showpacs,floatfix]{revtex4-1}
\usepackage{amsmath,amssymb,amsfonts,float,graphics,epsfig,epstopdf,color,verbatim,tabularx,bm,multirow,appendix}
\usepackage[utf8]{inputenc}
\usepackage[T1]{fontenc}
\usepackage{xcolor}

\usepackage{dsfont}
\usepackage{textcomp}
\usepackage{yfonts}
\usepackage{bm}
\usepackage{subfigure}
\usepackage{mathrsfs}
\usepackage{graphicx}
\usepackage{verbatim}
\usepackage{hyperref}
\usepackage{multirow}
\usepackage{physics}
\usepackage{braket}
\usepackage[normalem]{ulem}
\usepackage{tikz}
	\usetikzlibrary{calc}
	\usetikzlibrary{shapes.multipart}

\newcommand{\di}{\mathrm{d}}

\renewcommand{\ol}[1]{\overline{#1}}
\newcommand{\comments}[1]{}
\newcommand{\mb}[1]{\mathbf{#1}}

\def\Z{\mathbb{Z}}

\newcommand{\stkout}[1]{\ifmmode\text{\sout{\ensuremath{#1}}}\else\sout{#1}\fi}

\makeatletter
\def\l@subsubsection#1#2{}
\makeatother

\begin{document}

\title{Higher-form symmetry breaking at Ising transitions}

\author{Jiarui Zhao}
\affiliation{Department of Physics and HKU-UCAS Joint Institute of Theoretical and Computational Physics, The University of Hong Kong, Pokfulam Road, Hong Kong SAR, China}

\author{Zheng Yan}
\affiliation{Department of Physics and HKU-UCAS Joint Institute of Theoretical and Computational Physics, The University of Hong Kong, Pokfulam Road, Hong Kong SAR, China}
\affiliation{State Key Laboratory of Surface Physics and Department of Physics, Fudan University, Shanghai 200438, China}

\author{Meng Cheng}
\email{m.cheng@yale.edu}
\affiliation{Department of Physics, Yale University, New Haven, CT 06520-8120, U.S.A}

\author{Zi Yang Meng}
\email{zymeng@hku.hk}
\affiliation{Department of Physics and HKU-UCAS Joint Institute of Theoretical and Computational Physics, The University of Hong Kong, Pokfulam Road, Hong Kong SAR, China}
%\affiliation{Beijing National Laboratory for Condensed Matter Physics and Institute of Physics, Chinese Academy of Sciences, Beijing 100190, China}
%\affiliation{Songshan Lake Materials Laboratory, Dongguan, Guangdong 523808, China}

\begin{abstract}
In recent years, new phases of matter that are beyond the Landau paradigm of symmetry breaking are accumulating, and to catch up with this fast development, new notions of global symmetry are introduced. Among them, the higher-form symmetry, whose symmetry charges are spatially extended, can be used to describe topologically ordered phases as the spontaneous breaking of the symmetry, and consequently unify the unconventional and conventional phases under the same conceptual framework. However, such conceptual tools have not been put into quantitative test except for certain solvable models, therefore limiting its usage in the more generic quantum many-body systems. In this work, we study $\Z_2$ higher-form symmetry in a quantum Ising model, which is dual to the global (0-form) Ising symmetry.  We compute the expectation value of the Ising disorder operator, which is a non-local order parameter for the higher-form symmetry, analytically in free scalar theories and through unbiased quantum Monte Carlo simulations for the interacting fixed point in (2+1)d.  From the scaling form of this extended object, we confirm that the higher-form symmetry is indeed spontaneously broken inside the paramagnetic, or quantum disordered phase (in the Landau sense), but remains symmetric in the ferromagnetic/ordered phase. At the Ising critical point, we find that the disorder operator also obeys a ``perimeter'' law scaling with possibly multiplicative power-law corrections. We discuss examples where both the global 0-form symmetry and the dual higher-form symmetry are preserved, in systems with a codimension-1 manifold of gapless points in momentum space. These results provide non-trivial working examples of higher-form symmetry operators, including the first direct computation of one-form order parameter in an interacting conformal field theory, and open the avenue for their generic implementation in quantum many-body systems.

\end{abstract}
\date{\today}
\maketitle

\section{Introduction}
Global symmetries are instrumental in organizing our understanding of phases of matter. The celebrated Landau paradiagm classifies phases according to broken symmetries, which also determines the universality classes of transitions between phases.
Symmetry principles become even more powerful from the point of view of long wavelength, low-energy physics, as the renormalization group fixed points (i.e. IR) often embody more symmetries than the microscopic lattice model (i.e. UV), which is the phenomenon of emergent symmetry~\cite{Moessner2001,Isakov2003,Senthil2004,YYHe2016,NvsenMa2019}. A common example is the emergence of continuous space-time symmetries in the field-theoretical description of a continuous phase transition~\cite{altland2010condensed}. It is even plausible that a critical point is determined up to finite choices by its full emergent symmetry, which is the basic philosophy (or educated guess) behind the conformal bootstrap program~\cite{Poland2019}.

Modern developments in quantum many-body physics have significantly broadened the scope of quantum phases beyond the Landau classification~\cite{Wen2019}. For these exotic phases, more general notions of global symmetry are called for to completely characterize the phases and the associated phase transitions. Intuitively, these ``beyond Landau'' phases do not have local order parameters. Instead, non-local observables are often needed to characterize them. For a well-known example, confined and deconfined phases of a gauge theory are distinguished by the behavior of the expectation value of Wilson loop operators~\cite{fradkin2013field,Gregor2011}. To incorporate such extended observables into the symmetry framework, higher-form symmetries~\cite{Nussinov2006, Nussinov2009, Gaiotto_2015}, and more generally algebraic symmetries~\cite{ji2019categorical, kong2020algebraic} have been introduced. These are symmetries whose charged objects are spatially extended, e.g. strings and membranes. In other words, their symmetry transformations only act nontrivially on extended objects. Most notably, spontaneous breaking of such higher symmetries can lead to highly entangled phases, such as topological order~\cite{Gaiotto_2015}. Therefore, even though topologically ordered phases are often said to be beyond the Landau paradiagm, they can actually be understood within a similar conceptual framework once higher symmetries are included. In addition, just as the usual global symmetries, higher-form symmetries can have quantum anomalies~\cite{Gaiotto_2015}, which lead to strong non-perturbative constraints on low-energy dynamics~\cite{Gaiotto2017}.

In this work, we make use of the prototypical continuous quantum phase transition, the Ising transition, to elucidate the functionality of the higher-form symmetry. The motivation to re-examine the well-understood Ising transition is the following: in addition to the defining 0-form $\Z_2$ symmetry, the topological requirement that $\Z_2$ domain walls must be closed (in the absence of spatial boundary) can be equivalently formulated as having an unbreakable $\Z_2$ $(D-1)$-form symmetry, where $D$ is the spatial dimension. Gapped phase on either side of the transition spontaneously breaks one and only one of the two symmetries. Therefore to correctly determine the full emergent internal symmetry in the Ising CFT, the $\Z_2$ higher-form symmetry should be taken into account. For $D=2$, the $1$-form symmetry manifests more clearly in the dual formulation~\cite{Wegner1971}, namely as the confinement-deconfinement transition of a $\Z_2$ gauge theory, which will shed light on higher-form symmetry breaking transitions in a concrete setting.
%Recently, it was proposed that

A basic question about a global symmetry is whether it is broken spontaneously or not in the ground state. For clarity, let us focus on the $D=2$ case. It is well-known that the Ising symmetric, or ``quantum disordered'' phase, spontaneously breaks the higher-form symmetry, and the opposite in the Ising symmetry-breaking phase. The fate at the critical point remains unclear to date. To diagonose higher-form symmetry breaking, we compute the ground state expectation value of the ``order parameter'' for the higher-form symmetry -- commonly known as the disorder operator in literature~\cite{KadanoffPRB1971,Fradkin2016,XCWu2020,YCWang2021,XCWu2021}, which creates a domain wall in the Ising system. Spontaneous breaking of the $\Z_2$ $1$-form symmetry is signified by the perimeter law for the disorder operator. In the dual formulation, the corresponding object is the Wilson loop operator.  Through large-scale QMC simulations, we find numerically that at the transition, the disorder operator defined on a rectangular region scales as $l^s e^{-a_1l}$, where $l$ is the perimeter of the region, and $s>0$ is a universal constant. We thus conclude that the 1-form symmetry is spontaneously broken at the (2+1)d Ising transition, and it remains so in the disordered phase of the model. This is in stark contrast with the $D=1$ case, where the disorder operator has a power-law decay. 

To corroborate the numerical results, we consider generally disorder operator corresponding to a 0-form $\Z_2$ symmetry in a free scalar theory in $D$ dimensions, which is a stable fixed point for $D\geq 3$. We show that for the kind of $\Z_2$ symmetry in this case, the disorder operator can be related to the 2nd Renyi entropy. Therefore, the disorder operator also obeys a ``perimeter'' (i.e. volume of the boundary) scaling, with possibly multiplicative power-law correction. Whether the higher-form symmetry is broken or not is determined by the subleading power-law corrections. We also discuss other free theories, such as a Fermi liquid, where the decay of the disorder operator is in between the ``perimeter'' and the ``area'' laws, and therefore no higher-form symmetry breaking.

The rest of the paper is organized as follows. In Sec.~\ref{sec:ii} we review higher-form symmetry and its spontaneous breaking, and its relevancy in conventional phases. We also consider higher-form symmetry breaking in free and interacting conformal field theories. In Sec.~\ref{sec:iii} we specialize to the setting of quantum Ising model in (2+1)d and define the disorder operator. Sec.~\ref{sec:iv} presents the main numerical results from quantum Monte Carlo simulations, which reveal the key evidence of the 1-form symmetry breaking at the $(2+1)$d Ising transition. Sec.~\ref{sec:v} outlines a few immediate directions about the higher-form symmetry breaking and their measurements in unbiased numerical treatments in other quantum many-body systems.

\section{Generalized global symmetry}
\label{sec:ii}
Consider a quantum many-body system in $D$ spatial dimensions. Global symmetries are unitary transformations which commute with the Hamiltonian. Typically the symmetry transformation is defined over the entire system, and charges of the global symmetry are carried by particle-like objects.

An important generalization of global symmetry is the higher-form symmetry~\cite{Gaiotto_2015}. For an integer $p\geq 0$, $p$-form symmetry transformations act nontrivially on $p$-dimensional objects.  In other words, ``charges'' of $p$-form symmetry are carried by extended objects. In this language, the usual global symmetry is 0-form as the particle-like object is of 0-dimension. $p$-form symmetry transformations themselves are unitary operators supported on each codimension-$p$ (i.e. spatial dimension $(D-p)$) closed submanifold $M_{D-p}$. In particular, it means that there are infinitely many symmetry transformations in the thermodynamic limit. In this work we will only consider discrete, Abelian higher-form symmetry, so for each submanifold $M_{D-p}$ the associated unitary operators form a finite Abelian group $G$.
Physically, higher-form symmetry means that the certain $p$-dimensional objects are charged under the group $G$, and the quantum numbers they carry constrain the processes of creation, annihilation and splitting etc. In particular, these extended objects are ``unbreakable'', i.e. they are always closed and can not end on $(p-1)$-dimensional objects.

For a concrete example, let us consider (2+1)$d$ $\Z_2$ gauge theory definend on a square lattice. Each edge of the lattice is associated with a $\Z_2$ gauge field (i.e. a qubit), subject to the Gauss's law at each site $v$:
\begin{equation}
	\prod_{e\ni v}\tau_e^x=1.
	\label{eqn:gauss}
\end{equation}
Here $e$ runs over edges ending on $v$.

The divergence-free condition implies that there are no electric charges in the gauge theory. In other words, all $\Z_2$ electric field lines must form loops. An electric loop can be created by applying the following operator along any closed path $\gamma$ on the lattice:
\begin{equation}
	W_e(\gamma)=\prod_{e\in \gamma}\tau_e^z.
	\label{}
\end{equation}
The corresponding $\Z_2$ 1-form symmetry operator is defined as
\begin{equation}
	W_m(\gamma^\star)=\prod_{e\perp \gamma^\star} \tau_e^x
	\label{eqn:Wm}
\end{equation}
for any closed path $\gamma^\star$ on the dual lattice. Here the subscript $m$ in $W_m$ indicates that this is actually the string operator for $\Z_2$ flux excitations. In field theory parlance, $W_e$ is the Wilson operator  of the $\Z_2$ gauge theory, and $W_m$ is the corresponding Gukov-Witten operator~\cite{gukov2008rigid}. 

\begin{figure}[t]
	\centering
	\includegraphics[width=\columnwidth]{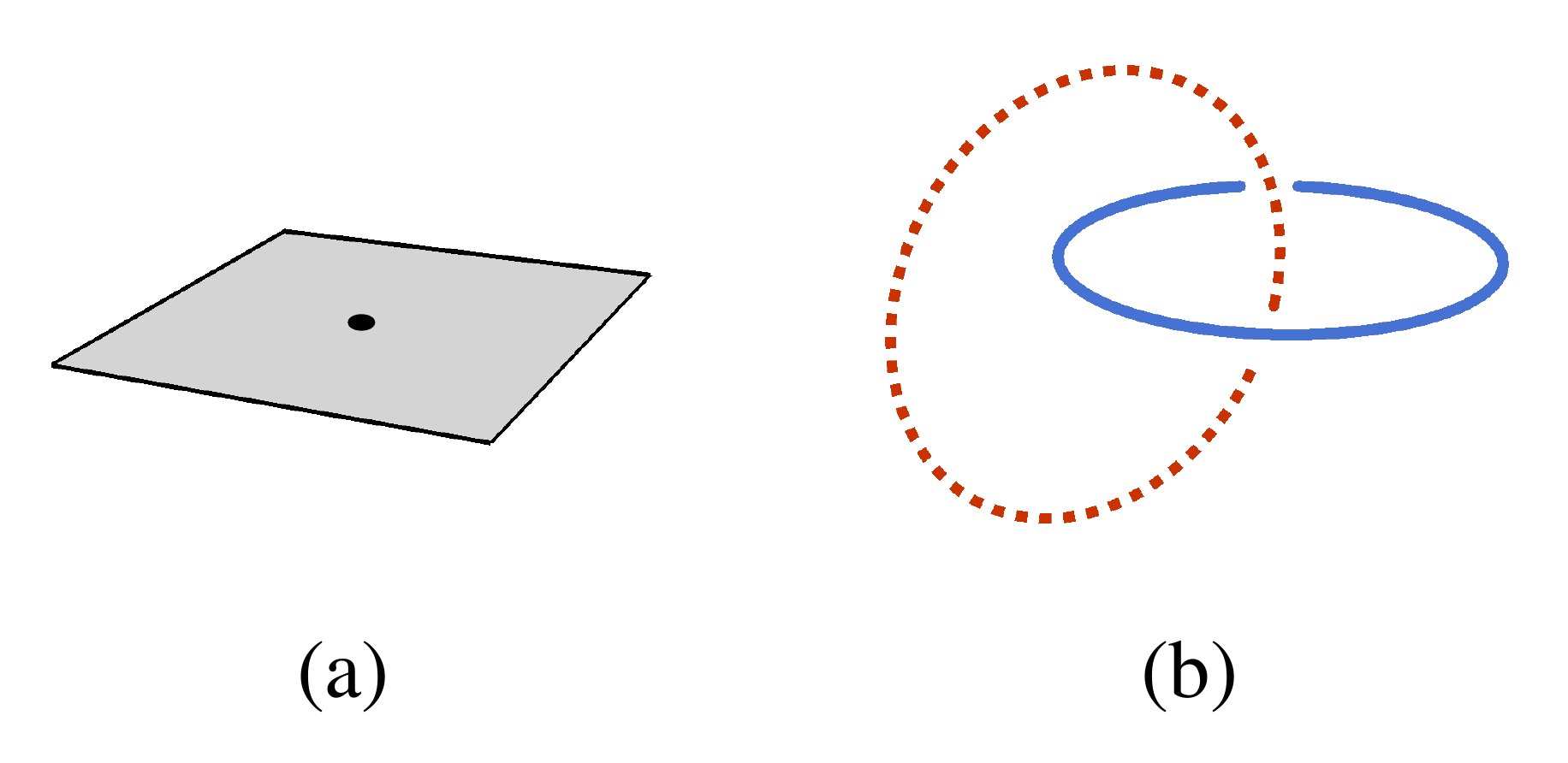}
	\caption{(a) 0-form symmetry charge is a point-like object, measured by the symmetry transformation defined on the entire system (i.e. at a fixed time slice) (b) 1-form symmetry charge is a loop (the solid line), measured by the symmetry transformation defined on a loop as well when the two are linked. }
	\label{fig:braiding}
\end{figure}

We notice that the $W_m(\gamma^\star)$ operator is in fact the product of Gauss's law term $\prod_{v\in e}\tau_e^x$ for all $v$ in the region enclosed by $\gamma^\star$. In other words, the smallest possible $\gamma^\star$ is a loop around one vertex $v$, and the fact tht $W_m(\gamma^\star)$ is conserved by the dynamics means that the gauge charge at site $v$ must be conserved (mod 2) as well. Therefore, the $\mathbb{Z}_2$ gauge theory with electric 1-form symmetry is one with completely static charges, including the case with no charges at all. For applications in relativistic quantum field theories, it is usually further required that the 1-form symmetry transformation is ``topological'', i.e. not affected by local deformation of the loop $\gamma^\star$, which is equivalent to the absence of gauge charge as given in Eq. \eqref{eqn:gauss}.

It is instructive to consider how the 1-form charge of an electric loop can be measured. This is most clearly done in space-time: to measure a $p$-dimensional charge, one ``wraps'' around the charge by a $(D-p)$-dimensional symmetry operator. Appying the symmetry transformation is equivalent to shrinking the symmetry operator, and in $(D+1)$ spacetime, because of the linking the two must collide, and the non-commutativity (e.g. between  $W_e$ and $W_m$) measures the charge value. We illustrate the process for $p=0$ (Fig. \ref{fig:braiding}(a)) and $p=1$ (Fig. \ref{fig:braiding}(b)), in three-dimensional space-time.

Now consider the following Hamiltonian of Ising gauge theory:
\begin{equation}
	H=-J\sum_e\tau_e^x - K\sum_p \prod_{e\in \partial p}\tau_e^z,
	\label{}
\end{equation}
where $J,K > 0$. When $J\ll K$, the ground state is in the deconfined phase, which can be viewed as an equal-weight superposition of all closed $\Z_2$ electric loops. In this phase, the $\Z_2$ 1-form symmetry is spontaneously broken.  When $J\gg K$, the ground state is a product state with $\tau_e^x=1$ everywhere, and the 1-form symmetry is preserved. This is the confined phase. Similar to the usual boson condensation, the expectation value of the electric loop creation operator $W_e(\gamma)$ can be used to characterize the 1-form symmetry breaking phase, which obeys perimeter law in the deconfined phase.

This example shows that higher-forms symmetry naturally arises in gauge theories. In condensed matter applications, gauge theories are usually emergent~\cite{Senthil2004,NvsenMa2018}, which means that dynamical gauge charges are inevitably present and the electric 1-form symmetry is explicitly broken. Even under such circumstances, at energy scales well below the electric charge gap, the theory still has an emergent 1-form symmetry~\cite{WenPRB2019}.

Let us now discuss more generally the spontaneous breaking of higher-form symmetry~\cite{Gaiotto_2015, Lake2018, Hofman2018}. We will assume that the symmetry group is discrete. For a $p$-form symmetry, a charged object is created by an extended operator $W(C)$ defined on a $p$-dimensional manifold $C$. When the symmetry is unbroken, we have
\begin{equation}
	\langle W(C)\rangle \sim e^{-t_{p+1} \mathrm{Area}(C)},
	\label{eq:eq5}
\end{equation}
where $\mathrm{Area}(C)$ is the volume of a minimal $(p+1)$-dimensional manifold whose boundary is $C$. $t_{p+1}$ can be understood as the ``tension'' of the $(p+1)$-dimensional manifold. This generalizes the exponential decay of charged local operator for the 0-form case. On the other hand, when the symmetry is spontaneously broken,
\begin{equation}
	\langle W(C)\rangle \sim e^{-t_p \mathrm{Perimeter}(C)},
	\label{}
\end{equation}
where $\mathrm{Perimeter}(C)$ denotes the ``volume'' of $C$ itself. Importantly the expectation value only depends locally on $C$, which is the analog of the factorization of the correlation function of local order parameter $\langle O(x)O^\dag(y)\rangle \approx \langle O(x)\rangle \langle O^\dag(y)\rangle$ for $0$-form symmetry. One can then redefine the operator $W(C)$ to remove the perimeter scaling and in that case $\langle W(C)\rangle$ would approach a constant in the limit of large $C$~\cite{HastingsPRB2005}. At critical point, however, subleading corrections become important, which will be examined below. 

The $\Z_2$ gauge theory is famously dual to a quantum Ising model~\cite{Kogut1979}. In fact, more generally, there is a duality transformation which relates a system with global $\Z_2$ 0-form symmetry (in the $\Z_2$ even sector) to one with global $\Z_2$ $(D-1)$-form symmetry, a generalization of the Kramers-Wannier duality in (1+1)d. 

Let us now review the duality in (2+1)d.   The dual Ising spins are defined on plaquettes, whose centers form the dual lattice. For a given edge $e$ of the original lattice, denote the two adjacent plaquettes by $p$ and $q$, as shown in the figure below:
\begin{center}
\begin{tikzpicture}
	\draw[thick] (-1.3, 0) -- (1.3, 0);
	\draw[thick] (-1.3, 1) -- (1.3, 1);
	\draw[thick] (-1.3,-1) -- (1.3,-1);
	\draw[thick, dashed] (-1.3,-0.5) -- (1.3,-0.5);
	\draw[thick, dashed] (-1.3,0.5) -- (1.3,0.5);
	\draw[thick] (-1, -1.3) -- (-1, 1.3);
	\draw[thick] (0, -1.3) -- (0, 1.3);
	\draw[thick] (1, -1.3) -- (1, 1.3);
	\draw[thick, dashed] (0.5, -1.3) -- (0.5, 1.3);
	\draw[thick, dashed] (-0.5, -1.3) -- (-0.5, 1.3);
	\filldraw (-0.5, -0.5) circle (1.5pt);
	\filldraw (0.5, -0.5) circle (1.5pt);
	\filldraw (-0.5, 0.5) circle (1.5pt);
	\filldraw (0.5, 0.5) circle (1.5pt);
	\node at (-0.35, 0.3) {$p$};
	\node at (0.61, 0.3) {$q$};
	\node at (0.13, 0.7) {$e$};
\end{tikzpicture}
\end{center}

The duality map is defined as follows:
\begin{equation}
	\sigma_{p}^z\sigma_{q}^z \leftrightarrow \tau_e^x, \sigma^x_{p} \leftrightarrow \prod_{e\in \partial p}\tau_e^z.
	\label{}
\end{equation}
Note that the expression automatically ensures $\prod_p\sigma^x_p=1$ in a closed system, so the dual spin system has a $\Z_2$ 0-form symmetry generated by $S=\prod_p \sigma_p^x$, and the map can only be done in the $\Z_2$ even sector with $S=1$~\footnote{In a sense the $\Z_2$ symmetry is gauged. In fact one way to derive the duality is to first gauge the $\Z_2$ symmetry and then perform gauge transformations to eliminate the Ising matter.}. Conversely, the mapping also implies $\prod_{v\in e}\tau_x^e=1$, and in fact $W_m(\gamma^\star)=1$ for any $\gamma^*$, i.e. the $\Z_2$ 1-form symmetry is strictly enforced.

In the dual model, the electric field line of the $\Z_2$ gauge theory becomes the domain walls separating regions with opposite Ising magnetizations. Therefore, a Wilson loop $W_e(\gamma)$ maps to
\begin{equation}
	X_M=\prod_{p\in M} \sigma_p^x,
	\label{eqn:Xdef}
\end{equation}
where $\partial M=\gamma$, i.e. $M$ is the region enclosed by $\gamma$. Physically $X_M$ flips all the Ising spins in the region $M$, thus creating a domain wall along the boundary $\gamma$. It is called the disorder operator for the Ising system, which will be the focus of our study below.

Under the duality map, the Hamiltonian becomes
\begin{equation}
	H=-J\sum_{\langle pq\rangle}\sigma_p^z\sigma_q^z - K\sum_p \sigma_p^x.
	\label{eqn:TFI}
\end{equation}

The phases of the gauge theory can be readily understood in the dual representation. For $K\gg J$, the $\Z_2$ gauge theory is in the deconfined phase, which means that the ground state contains arbitrarily large electric loops. For the dual Ising model, the ground state is disordered, with all $\sigma_p^x=1$. If we work in the $\sigma^z$ eigenbasis (which is natural to discuss symmetry breaking), the ground state wavefunction is given by
\begin{equation}
	\ket{\psi_{K=\infty}}\propto \prod_p \frac{1+\sigma_p^x}{2}\ket{\uparrow\uparrow\cdots\uparrow}.
	\label{}
\end{equation}
Namely we pick any basis state and apply the ground state projector. Expanding out the projector, one can see that the wavefunction is an equal superposition of all domain wall configurations, i.e. a condensation of domain walls. Since the domain walls carry $\Z_2$ 1-form charges, the condensation breaks the 1-form symmetry spontaneously, much like the Bose condensation spontaneously breaks the conservation of particle numbers

In the other limit $K\ll J$, the gauge theory is confined. Correspondingly, the dual Ising model is in the ferromagnetically ordered phase: there are two degenerate ground states $\ket{\uparrow\cdots\uparrow}$ and $\ket{\downarrow\cdots\downarrow}$. There are no domain walls at all in the limit $K\rightarrow 0$. When a small but finite $K/J$ is turned on, quantum fluctuations create domain walls on top of the fully polarized ground states, but these domain walls are small and sparse.

\subsection{Non-invertible anomaly and gapless states}
A notable feature of the duality map is that on either side, only one of two symmetries, the $\Z_2$ 0-form and the $\Z_2$ 1-form symmetries, is faithfully represented (in the sense that the symmetry transformation is implemented by a nontrivial operator, even though the duality is supposed to work only in the symmetric sector). The other symmetry transformation is mapped to the identity at the operator level. Physically, only one of them is an explicit global symmetry, while the other one appears as a global constraint (e.g. on the Ising side, domain walls of the 0-form global symmetry are codimension-1 closed manifolds, which is the manifestation that they are charged under a $(D-1)$-form symmetry). 

A closely related fact is that the ordered phase for one symmetry is necessarily the disordered phase of the other, and any non-degenerate gapped phase must break one and only one of the two symmetries. This has been proven rigorously in one spatial dimension~\cite{Levin2019}, and is believed to hold in general dimensions as well.

It is clear from these results that these two symmetries can not be considered as completely independent. Recently, Ref. [\onlinecite{JiPRR2019}] proposed that the precise relation between the two dual symmetries is captured by the notion of a non-invertible quantum anomaly. Intuively, the meaning of the non-invertible anomaly in the context of the $\Z_2$ Ising model can be understood as follows: the charge of the $\Z_2$ 0-form symmetry is an Ising spin flip, while the charge of the $\Z_2$ 1-form symmetry is an Ising domain wall. These two objects have nontrivial mutual ``braiding'', in the sense that when an Ising charge is moved across a domain wall, it picks up a minus sign due to the Ising symmetry transformation applied to one side of the domain wall. In other words, the charge of the 1-form symmetry is actually a flux loop of the 0-form symmetry. Ref. [\onlinecite{JiPRR2019}] suggested that two symmetries whose charged objects braid nontrivially with each other can not be realized faithfully in a local Hilbert space. If locality is insisted, then the only option is to realize the $D$ spatial dimensional system as the boundary of a $\Z_2$ toric code model in $(D+1)$ spatial dimension. In this case, the charged objects are in fact bulk topological excitations brought to the boundary. The nontrivial braiding statistics between the two kinds of charges reflects the topological order in the bulk. Such an anomaly is fundamentally different from more familiar 't Hooft anomaly realized on the boundary of a symmetry-protected topological phase (which is an invertible state). We refer to Ref. [\onlinecite{JiPRR2019}] for more thorough discussions of the non-invertible anomaly.

 Since any gapped state must break one of the two symmetries, it is a very natural question to ask whether there are gapless states that preserve both symmetries.  An obvious candidate for such a gapless state is the symmetry-breaking continuous transition.   At the transition, the two-point correlation function of the Ising order parameter decays algebraically with the distance, implying that the $\Z_2$ 0-form symmetry is indeed unbroken. For the dual $(D-1)$-form symmetry,  the Kramers-Wannier duality maps the disorder operator, which is a string operator in the Ising basis, to the two-point correlator of the Ising order parameter. Therefore the expectation value of the disorder operator also exhibits power-law correlation, and the dual $0$-form symmetry is preserved. Therefore the Ising conformal field theory in (1+1)d indeed provides an example of symmetric gapless state with non-invertible anomaly~\cite{JiPRR2019}. But for the case of $D>1$, the situation is far from clear and that is what we will address in this paper. First we analyze the expectation value of the disorder operator in a free field theory.

\subsection{Scaling of disorder operator in field theory}
We now discuss the scaling form of the disorder operator at or near the critical point from a field-theoretical point of view.  The natural starting point is the Gaussian fixed point, i.e. a free scalar theory, described by the following Hamiltonian
\begin{equation}
	{H}[\phi]=\int\di^D\mathbf{r}\,\left[\frac{\pi^2}{2}+\frac{1}{2}(\nabla \phi)^2\right].
	\label{eqn:freeboson1}
\end{equation}
The real scalar $\phi$ can be thought of as the coarse grained Ising order parameter, and $\pi$ is the conjugate momentum of the real scalar $\phi$. The $\Z_2$ symmetry acts as $\phi\rightarrow -\phi$. The disorder operator $X_M$ is basically defined as the continuum version of Eq. \eqref{eqn:Xdef}, where the $\Z_2$ symmetry is applied to a finite region $M$.

Interestingly, for the free theory the expectation value of the disorder operator can be related to another well-studied quantity, the 2nd Renyi entanglement entropy $S_2$. More precisely, for a region $M$, we have
\begin{equation}
	e^{-S_2(M)}=\langle X_M\rangle.
	\label{eqn:S2=X}
\end{equation}
Here $S_2(M)$ is the 2nd Renyi entropy of the region $M$.

To see why this is the case, recall that the 2nd Renyi entropy $S_2$ for a region $M$ of a quantum state $\ket{\Psi}$ is given by
\begin{equation}
	e^{-S_2(M)}=\Tr \rho_M^2,
	\label{}
\end{equation}
where $\rho_M$ is the reduced density matrix for the region $M$, obtained from tracing out the degrees of freedom in the complement $\ol{M}$: $\rho_M= \Tr_{\ol{M}} \ket{\Psi}\bra{\Psi}$. In the following we denote the ground wave functional of the state $\ket{\Psi}$ by $\Psi(\phi)$:
\begin{equation}
	\ket{\Psi}=\int D\phi\, \Psi(\phi)\ket{\phi}.
	\label{eqn:wfn}
\end{equation}

The Renyi entropy can be calculated with a replica trick, which we now review in the Hamiltonian formalism. Consider two identical copies of the system, in the state $\ket{\Psi}\otimes\ket{\Psi}$. In the field theory example, the fields in the two copies are denoted by $\phi^{(1)}$ and $\phi^{(2)}$, respectively. We denote the basis state with a given field configuration $\phi^{(i)}$ in the $i$-th copy by $\ket{\phi^{(i)}_M,\phi^{(i)}_{\ol{M}}}$, where $\phi^{(i)}_M$ is the field configuration restricted to $M$ and similarly $\phi^{(i)}_{\ol{M}}$ for the complement of $M$. Since the two copies are completely identical, there is a swap symmetry $R$ acting between the two copies $R: \phi^{(1)}\leftrightarrow \phi^{(2)}$. $R_M$ then swaps the field configurations only within the region $M$:
\begin{equation}
	R_M\ket{\phi^{(1)}_M, \phi^{(1)}_{\ol{M}}}\otimes\ket{\phi^{(2)}_M, \phi^{(2)}_{\ol{M}}}=
	\ket{\phi^{(2)}_M, \phi^{(1)}_{\ol{M}}}\otimes\ket{\phi^{(1)}_M, \phi^{(2)}_{\ol{M}}}.
	\label{}
\end{equation}

The expectation of $R_M$ on the replicated ground state $\ket{\Psi}\otimes\ket{\Psi}$ is then given by
\begin{equation}
	\begin{split}
	(\bra{\Psi}&\otimes\bra{\Psi})R_M(\ket{\Psi}\otimes\ket{\Psi})\\
	&=\int \prod_{i=1,2}D\phi^{(i)}_M D\phi^{(i)}_{\ol{M}}\,\Psi(\phi^{(1)}_M,\phi^{(1)}_{\ol{M}})\Psi^*(\phi^{(2)}_M,\phi^{(1)}_{\ol{M}})\\
	&\quad\quad\Psi(\phi^{(2)}_M,\phi^{(2)}_{\ol{M}})\Psi^*(\phi^{(1)}_M,\phi^{(2)}_{\ol{M}})\\
	&=\int D\phi^{(1)}_M D\phi^{(2)}_{M}\,\rho_M(\phi^{(1)}_M, \phi^{(2)}_M)\rho_M(\phi^{(2)}_M, \phi^{(1)}_M)\\
	&=\Tr \rho_M^2.
	\end{split}
	\label{}
\end{equation}
Therefore the Renyi entropy is the expectation value of the disorder operator for the replica symmetry.

For a free theory,  we rotate the basis to $\phi_\pm = \frac{1}{\sqrt{2}}(\phi^{(1)}\pm \phi^{(2)})$. In the new basis, the swap symmetry operator becomes:
\begin{equation}
	R:\phi_\pm \rightarrow \pm \phi_\pm.
	\label{}
\end{equation}
It is straightforward to check that the Hamiltonian of the replica takes essentially the same form in the new basis:
\begin{equation}
	H[\phi^{(1)}]+H[\phi^{(2)}]=H[\phi_+]+H[\phi_-].
	\label{}
\end{equation}
The ground state again is factorized: $\ket{\Psi}\otimes\ket{\Psi}=\ket{\Psi}_+\otimes\ket{\Psi}_-$, where $\ket{\Psi}_\pm$ is the state of the $\phi_\pm$ field, with the same wave functional as $\phi$: $\braket{\phi_\pm |\Psi}_\pm = \Psi(\phi_\pm)$ as defined in Eq. \eqref{eqn:wfn}.

We can now compute the expectation value of $R_M$:
\begin{equation}
	(\bra{\Psi}_+\otimes\bra{\Psi}_-)R_M(\ket{\Psi}_+\otimes\ket{\Psi}_-) =\braket{X_M}.
	\label{}
\end{equation}
where we used the fact that $R$ acts as the identity on $\phi_+$. For $\phi_-$, $R_M$ is nothing but the disorder operator $X_M$. 

The 2nd Renyi entropy of a free scalar has been well-studied~\cite{Casini2006,Casini2009, Dowker2015, ElvangPLB2015, Helmes2016, Bueno2019, Berthiere2018} and we summarize the results below.

It is important to distinguish the case where the boundary is smooth and those with sharp corners on the boundary.

First consider a smooth boundary. For a sphere of radius $R$,  in $D=1,2,3$ we have:
\begin{equation}
	S_2=
	\begin{cases}
		\frac{1}{6}\ln R & D=1\\
		a_1\frac{R}{\epsilon} -\gamma & D=2\\
		a_2\left(\frac{R}{\epsilon}\right)^2-\frac{1}{192}\ln \frac{R}{\epsilon} & D=3
	\end{cases}.
	\label{eqn:S2scaling}
\end{equation}
Here $\epsilon$ is a short-distance cutoff, e.g. the lattice spacing, $a_1, a_2$ non-universal coefficients and $\gamma$ a universal constant.   For a more general smooth entangling boundary, in 2D the same form holds although the constant correction $\gamma$ depends on shape of the region. In 3D, it is known that the coefficient of the logarithmic divergent part of the Renyi entropy can be determined entirely from the local geometric data (e.g. curvature) of the surface in a general CFT~\cite{Solodukhin2008, Fursaev2012}.  

If the boundary has sharp corners then there are additional divergent terms in the entropy. The prototypical case is $D=2$ when the entangling region has sharp corners. In that case
\begin{equation}
	S_2=a_1\frac{l}{\epsilon}-s\ln \frac{l}{\epsilon},
	\label{eq:eq16}
\end{equation}
where $l$ is the perimeter of the entangling region and $s$ is an universal function that only depends on the opening angles of the corners.
For real free scalar, the coefficient of the logarithmic correction is $s\approx 0.0260$ for a square region (so four $\pi/2$ corners, as those in Fig.~\ref{fig:fig1})~\cite{Casini2009, Helmes2016}.  

Qualitatively, it is important that for $D=2,3$ the leading term in $S_2$ always obeys an ``perimeter'' law, i.e. it only depends on the ``area'' (length in 2D) of the entangling boundary. If instead we view $S_2$ as the disorder operator for the $\Z_2$ replica symmetry, the non-universal, cutoff-dependent perimeter term can be removed by redefining the disorder operator locally along the boundary, and the remaining term is universal. For $D=2$, the subleading term is either a \emph{negative} constant when the boundary is smooth, or a $\ln l$ correction with a \emph{negative} coefficient. So according to the relation Eq.~\eqref{eqn:S2=X}, the disorder parameter $\braket{X_M}$, after renormalizing away the perimeter term, does not decrease with the size of $M$, and therefore the corresponding $(D-1)$-form symmetry is spontaneously broken. This is consistent with the fact that the replica symmetry itself must be preserved as there is no coupling between the two copies.

Although the free Gaussian theory is unstable against quartic interactions below the upper critical dimension, and the actual critical theory is the interacting Wilson-Fisher fixed point, results from the free theory can still provide useful insights.  It is well-known that for $D=1$, for $M$ an interval of length $R$ the disorder operator $\braket{X_M}\sim R^{-1/4}$, the same power-law decay as that of the Ising order parameter due to Kramers-Wannier duality. For $D=2$, we will resort to numerical simulations below to address the question.

 Notice that the relation between $\langle X\rangle$ and $S_2$ essentially holds for all free theories, including free fermions. For example, the disorder operator associated with the fermion parity symmetry is also equal to $S_2$. Interestingly, for a Fermi liquid, it is well-known that $\ln \langle X\rangle = - S_2\sim -l^{D-1}\ln l$~\cite{Gioev2006, Wolf2006}, where here $l$ is the linear size of the region.  This is an example of a gapless state where the $(D-1)$-form symmetry is preserved. Similar results hold for non-interacting bosonic systems with ``Bose surface''~\cite{LaiPRL2013}, an example of which in 2D is given by the exciton Bose liquid~\cite{ParamekantiPRB2002, TayPRL2010}:
 \begin{equation}
	 H=\int\di^2\mathbf{r}\,\left[\frac{\pi^2}{2}+\kappa (\partial_x\partial_y \phi)^2\right].
	 \label{}
 \end{equation}
\newline
In other words, to preserve both the $0$-form symmetry and the dual $(D-1)$-form symmetry, it is necessary to have a surface of gapless modes in the momentum space.

While analytical results discussed in this work are limited to free theories,  we conjecture that similar scaling relations hold for interacting CFTs as well. To see why this is plausible, we notice that the entanglement Hamiltonian of a CFT is algebraically ``localized'' near the boundary of the subsystem~\cite{Casini2011}, which suggests that even for a non-local observable, such as the disorder operator, the major contribution is expected to come from the boundary, and hence a perimeter law scaling. We leave a more systematic study along these lines for future work. In Sec. \ref{sec:iv} we numerically confirm our conjecture for the Ising CFT in (2+1)d.

We now briefly discuss what happens if a small mass is turned on in Eq. \eqref{eqn:freeboson1}. Suppose we are in a gapped phase, and denote by $\xi$ the correlation length. In general, we expect that $S_2$ obeys an perimeter scaling in the gapped phase, namely the leading term in $S_2$ is given by $a\frac{R}{\epsilon}$. In 2D for a disk entangling region of radius $R$, we have~\cite{Metlitski2009EE}
\begin{equation}
	S_2= a_{c}\frac{R}{\xi} + f\left( \frac{R}{\xi} \right).
	\label{eq:eq17}
\end{equation}
Here $a_{c}$ is the value of $a$ at the critical point (which was denoted by $a_1$ in Eq. \eqref{eqn:S2scaling}).  The function $f(x)$ satisfies
\begin{equation}
	f(x)\rightarrow
	\begin{cases}
		rx & x\rightarrow \infty\\
		-\gamma_c & x\rightarrow 0
	\end{cases}.
	\label{}
\end{equation}
Here $r$ is an universal constant (once the definition of $\xi$ is fixed). Suppose the transition is tuned by an external parameter $g$ and the critical point is reached at $g_c$. Since $\xi\sim (g-g_c)^{-\nu}$ where $\nu$ is the correlation length exponent, one finds that
\begin{equation}
	a-a_c \sim (g-g_c)^{\nu},
	\label{eq:eq19}
\end{equation}

\section{Order and disorder in Ising spin models}
\label{sec:iii}

In the following we study $1$-form symmetry breaking in the transverse field Ising (TFI) model which gives rise to the $(2+1)$d Ising transition. We have reviewed the connection with the $\Z_2$ gauge thory in Sec. \ref{sec:ii}, as well as the 1-form symmetry in the Ising spin system.  We will now focus more on the quantitative aspects of the TFI model. Even though the TFI model and the $\Z_2$ lattice gauge theory are equivalent by the duality map,  we choose to work with the TFI model here because the numerical simulation is more straightforward.

 We will now consider a square lattice with one Ising spin per site, and the global Ising symmetry is generated by $S=\prod_\mb{r}\sigma_\mb{r}^x$.  There are, generally speaking, two phases: a ``disordered'' phase, where the Ising symmetry is preserved by the ground state~\footnote{We note that there are in fact two distinct types of Ising-disordered phases in 2D, one trivial paramagnet and the other one a nontrivial Ising symmetry-protected topological phase.}, and an ordered phase where the ground states spontaneously break the symmetry. They are separated by a quantum phase transition, described by a conformal field theory with $\Z_2$ symmetry. It is well-understood how to characterize the Ising symmetry breaking (and its absence) in the three cases: consider the two-point correlation function of the order parameter $\sigma^z_\mb{r}$. The asympotic forms of the correlation function $\langle \sigma_\mb{r}^z\sigma_\mb{r'}^z\rangle$ for large $|\mb{r}-\mb{r}'|$ distinguish the three cases:
\begin{equation}
	\langle \sigma_\mb{r}^z\sigma_\mb{r'}^z\rangle\sim
	\begin{cases}
		e^{-\frac{|\mb{r}-\mb{r}'|}{\xi}} & \text{disordered}\\
		\frac{1}{|\mb{r}-\mb{r}'|^{2\Delta}} & \text{critical}\\
		\text{const.} & \text{ordered}
	\end{cases}.
	\label{}
\end{equation}
In both the disordered phase and the quantum critical point, the Ising symmetry is preserved because of the absence of long-range order. The prototypical lattice model that displays all these features is the TFI model defined on a square lattice:
\begin{equation}
	H=-\sum_{\langle \mb{r}\mb{r'}\rangle}\sigma_\mb{r}^z\sigma_{\mb{r}'}^z - h\sum_\mb{r} \sigma_\mb{r}^x, h\geq 0.
	\label{eq:eq6}
\end{equation}
Note that this is the same as Eq. \eqref{eqn:TFI}, but we have set $J=1$ and renamed $K$ by $h$, to align with the standard convention in literature.  The model is in the ordered (disordered) phase for $h\ll 1$ ($h\gg 1$). The precise location of the critical point varies with dimension, $h_c=1$ in $D=1$ and $h_c=3.044$ in $D=2$~\cite{Bloete2002,ZiHongLiu2019}.

\begin{figure}
	\centering
	\includegraphics[width=0.72\linewidth]{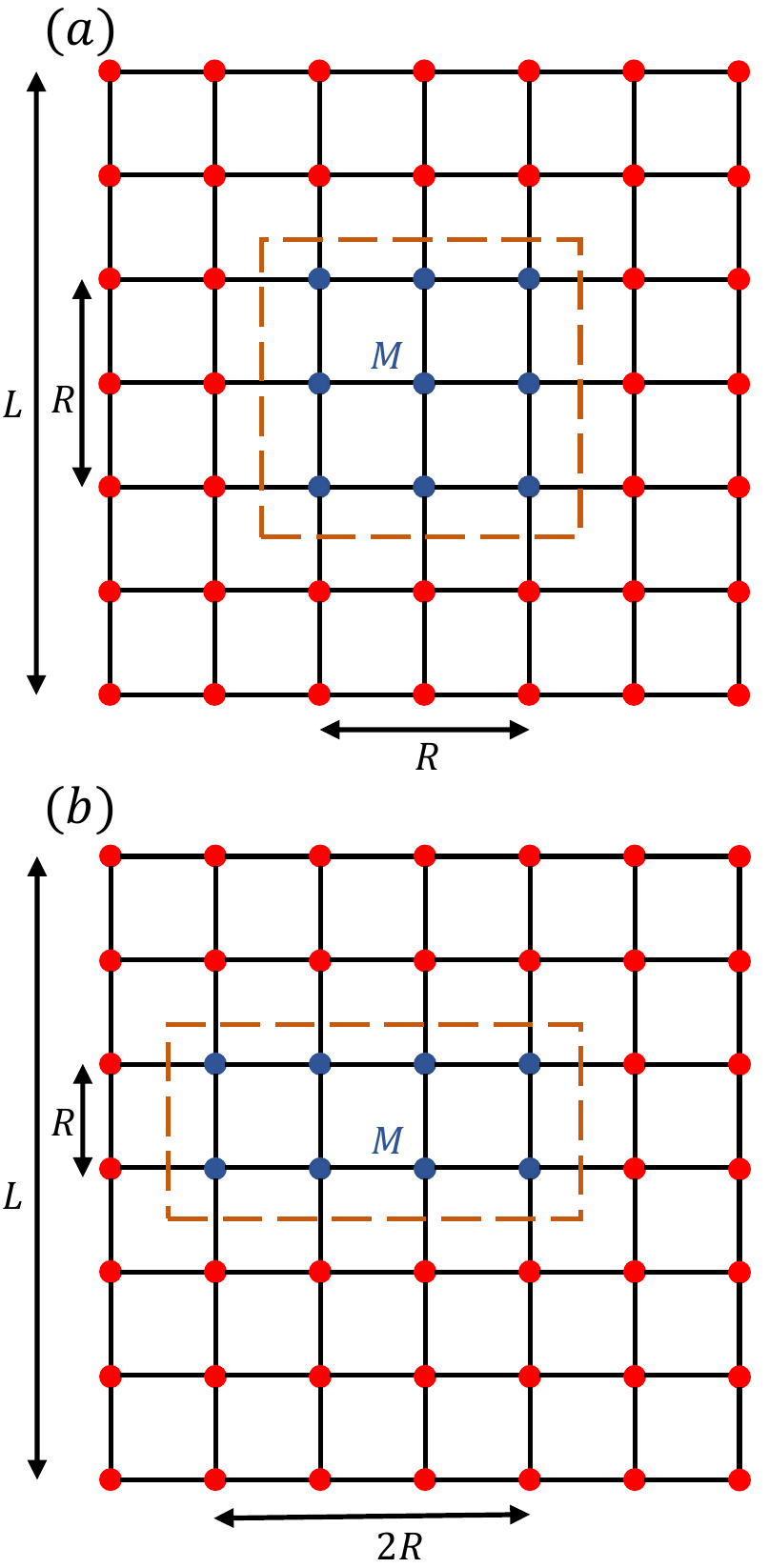}
	\caption{Disorder operator $X$ applied on regions with different shapes: (a) $M$ is a square region with size $R\times R$ and perimeter $l$. (b) $M$ is a rectangular region with size $R\times 2R$.}
	\label{fig:fig1}
\end{figure}

 We will be interested in the disorder operator:
\begin{equation}
	X_{ M}=\prod_{\mb{r}\in M}\sigma_\mb{r}^x,
	\label{eq:eq8}
\end{equation}
where $M$ is a rectangle region in the lattice, illustrated in Fig.~\ref{fig:fig1}.
In Ref.~\onlinecite{ji2019categorical} this operator is called the patch symmetry operator.

 When $X_M$ is applied to e.g. $\ket{\uparrow\cdots\uparrow}$, a domain wall is created along the boundary of the region $M$. These operators are charged under the dual $\Z_2$ 1-form symmetry. One can easily see that $\bra{\psi_{h=\infty}}X_M\ket{\psi_{h=\infty}}=1$, and $\bra{\psi_{h=0}}X_M\ket{\psi_{h=0}}=0$. More generally,
\begin{equation}
	\bra{\psi}X_M\ket{\psi} \sim
	\begin{cases}
		e^{-al_M} & h>h_c\\
		e^{-bA_M} & h<h_c
	\end{cases}.
	\label{eq:eq9}
\end{equation}
when $M$ is sufficiently large compared to the correlation length.
Here $l$ is the perimeter of the boundary of $M$, and $A$ is the area of $M$.  The coefficients $a$ and $b$ can be computed perturbatively in the limit of large and small $h$. In 2D, take $M$ to be a square of perimeter $l$, so Perimeter$(M)=l$ and the Area$(M)=l^2/16$. We can find that for large $l$:
\begin{equation}
	 -\ln\langle X\rangle=
	 \begin{cases}
		 \frac{l}{8h^2} & h\gg h_c\\
		 \frac{1}{4}|\ln h| l^2 & h\ll h_c
	 \end{cases}.
	\label{eq:eq10}
\end{equation}

\section{Numerical Simulations}
\label{sec:iv}
In this section we study the disorder operator in the (2+1)d TFI model.
 We employ the Stochastic Series Expansion (SSE) quantum Monte Carlo method~\cite{sandvikTFIM,OFS2002,OFS2003,YDLiao2021} to simulate the Hamiltonian in Eq.~\eqref{eq:eq6}. In particular, to be able to directly access the disorder operator in Eq.~\eqref{eq:eq8}, instead of implementing the algorithm in the conventional $\sigma^{z}$ basis we choose to work in the $\sigma^{x}$ basis and construct the highly efficient directed loop algorithm therein~\cite{OFS2002}. The implementation details of the SSE-QMC algorithm are given in Appendix~\ref{sec:app2}.

In our numerical simulations, we choose $M$ to be a rectangular region of size $R_1\times R_2$ (i.e. the region contains $R_1R_2$ sites), and denote the perimeter $l=2(R_1+R_2)$. As shown in Fig.~\ref{fig:fig1} (a) and (b), for finite-size studies, we fix the aspect ratio $R_2/R_1=1$ of square shape and $2$ of rectangle shape. The linear system size of the lattice is $L$ and at the critical point we scale the inverse temperature $\beta=1/T \sim L$ to access the thermodynamic limit.

\begin{figure}[htp!]
	\centering
	\includegraphics[width=\columnwidth]{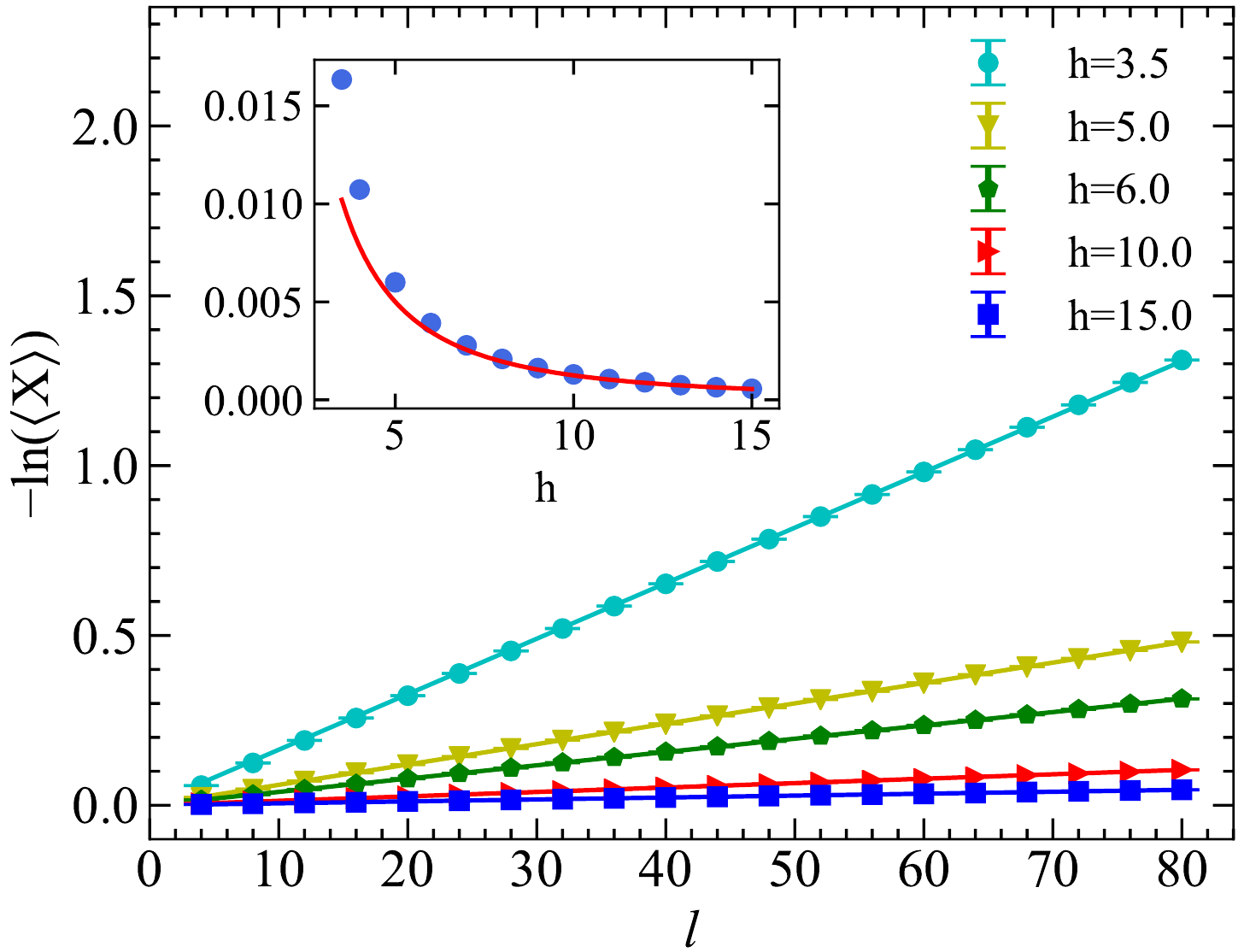}
	\caption{$-\ln(\langle X \rangle)$ versus $l$ at $L=32$ for different $h$ in disordered phases. We use the straight line to fit the data of different external fields and put the obtained slopes in the inset, one sees that as $h\gg h_c$, the fitted slopes (blue circles) approach the predicted relation $y=\frac{1}{8h^{2}}$ (red line). The fitting errors are negligible compared to the circle size.}
	\label{fig:fig2}
\end{figure}

\subsection{Disordered phase $h>h_c$}

First we present results in the disordered phase $h>h_c$. As shown in Eq.~\eqref{eq:eq9}, we expect that the disorder operator obeys a perimeter law scaling, and for $h\gg h_c$ the coefficient is given in Eq.~\eqref{eq:eq10}.

Fig.~\ref{fig:fig2} shows the QMC-obtained $\ln\langle X_M \rangle$ as a function of $l$ for different values of $h$. The temperature is taken to be $\beta=10$, and we have checked that the results already converge for this value of $\beta$. We observe a clear linear scaling, and the inset shows that for large field $h\gg h_c$, the slopes of the $\ln\langle X_M \rangle$ are indeed given by $1/8h^{2}$ asympototically.

\begin{figure}[htp!]
	\centering
	\includegraphics[width=\columnwidth]{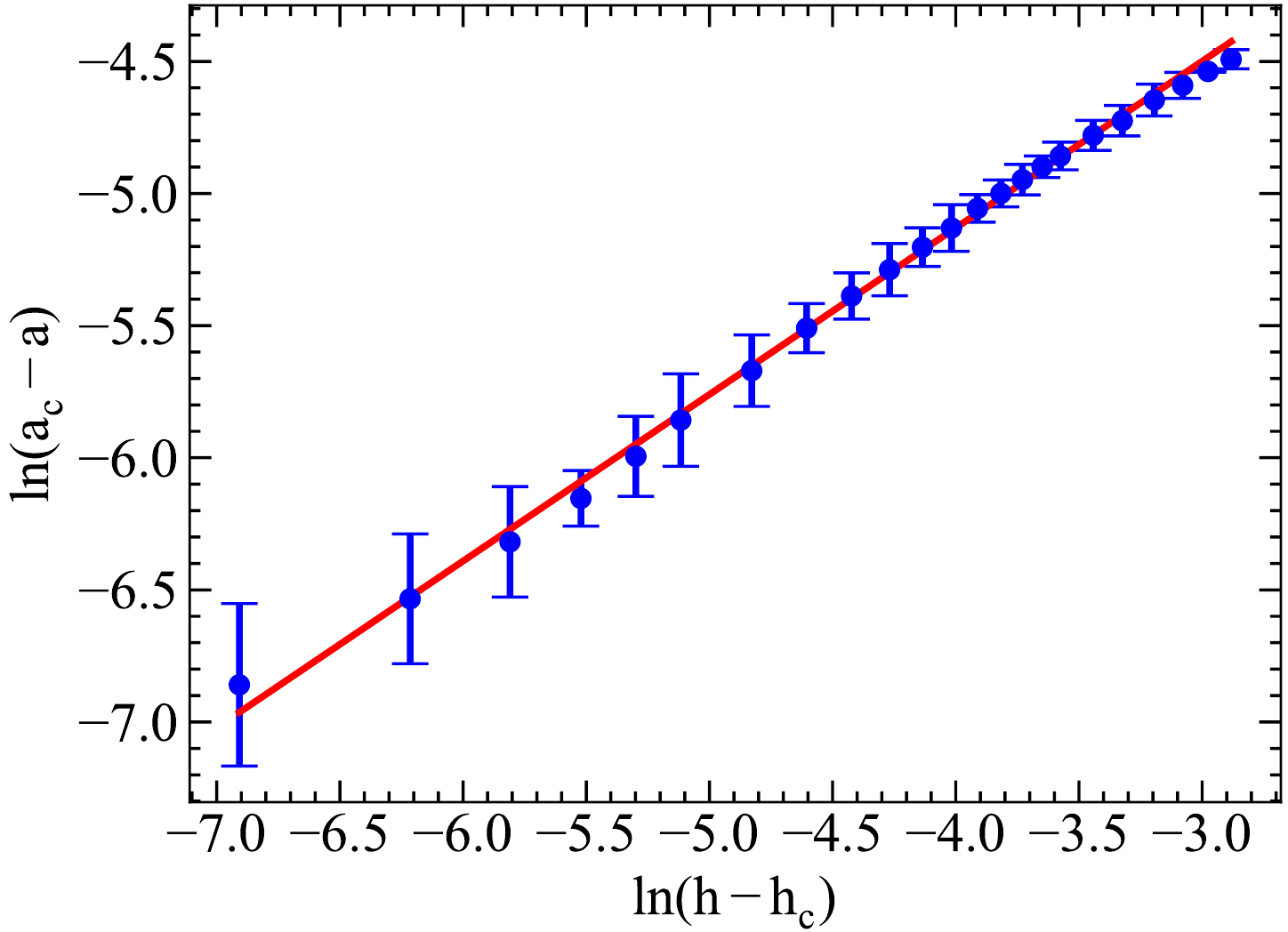}
	\caption{$\ln(a_{c} - a)$ versus $\ln(h-h_{c})$ in the disordered phase for $L=24$ when $h$ is approaching the critical point. The fitted slope (red line) is $0.63\pm 0.02$, consistent with the correlation length exponent of the $(2+1)$d Ising transition, as expected in Eq.~\eqref{eq:eq19}.}
	\label{fig:fig3}
\end{figure}

Now we consider the other limit, when $h$ is approaching the critical point $h_c$ from the disordered side. To test the scaling given in Eq.~\eqref{eq:eq19}, we measure the disorder operator and find the slope $a$ by a linear fit. Fig.~\ref{fig:fig3} shows $a_c-a$ as a funtion of $h-h_c$ in a log-log plot. A clear power law manifests in the data, and the exponent is found to be $\nu=0.63(2)$. Considering the finite-size effect, the result agrees very well with the 3D Ising correlation length exponent.

\subsection{Critical point $h=h_c$}
 The central question to be addressed is whether the $\Z_2$ 1-form symmetry is spontanously broken at the critical point. To this end, we measure the disorder operator $\langle X \rangle$ at $h=h_c$ and scale the inverse temperature $\beta=L$ in these simulations. We have also checked that finite-$\beta$ effect is negligible in our calculations.

\begin{figure}[htp!]
	\centering
	\includegraphics[width=\columnwidth]{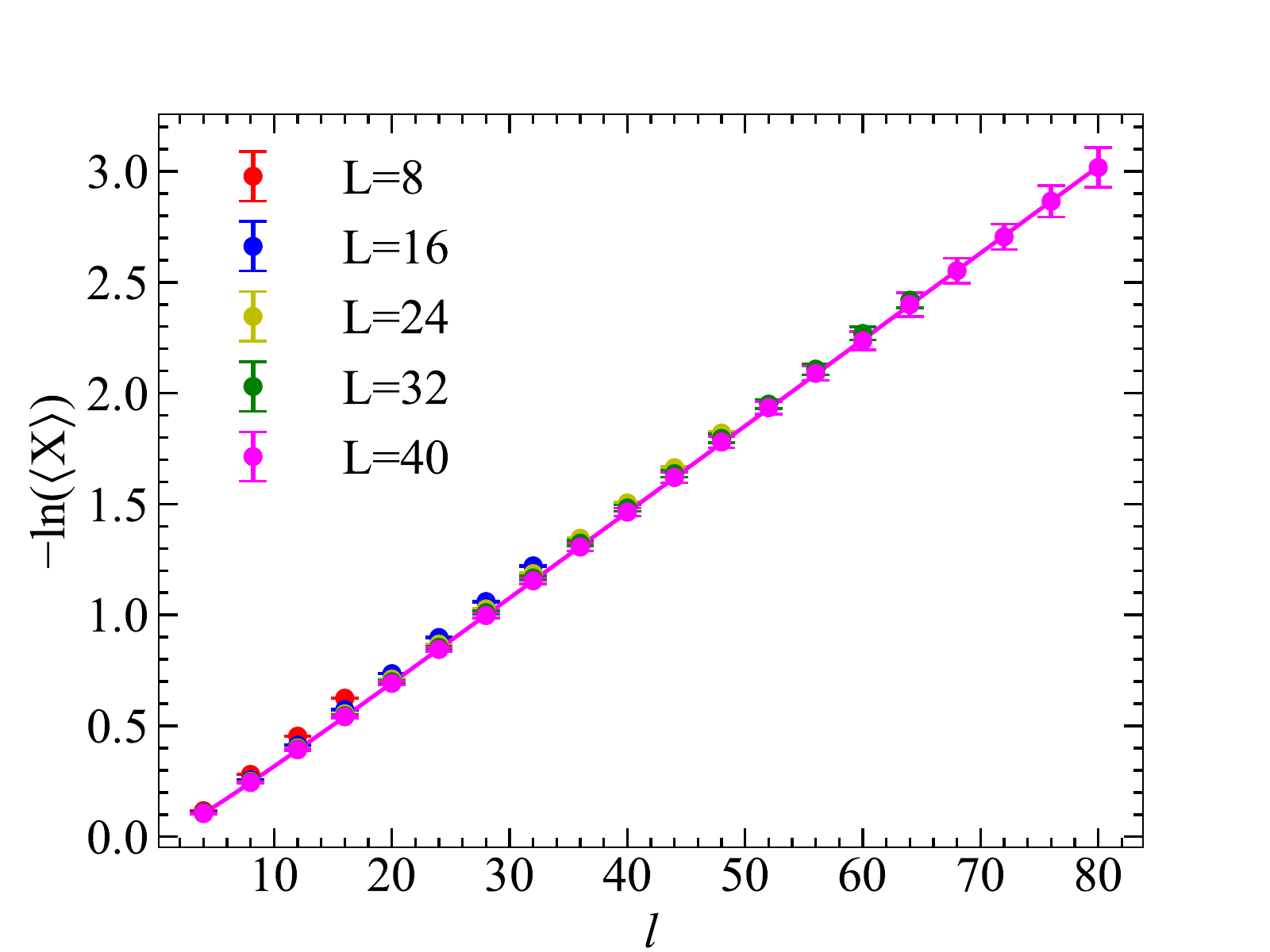}
	\caption{$-\ln(\langle X \rangle)$ versus $l$ at the critical point. We use the relation of Eq.~\eqref{eq:eq16} to fit the data and the fitted curve of the data upto $L=40$ is $-\ln(\langle X \rangle)=(0.0394 \pm 0.0004)l-(0.0267 \pm 0.005)\ln(l)-(0.0158 \pm 0.008)$.}
	\label{fig:fig4}
\end{figure}

Fig.~\ref{fig:fig4} shows $\ln\langle X_{M} \rangle$ as a funtion of the perimeter $l$, where $M$ is taken to be a square region,  as illustrated in Fig.~\ref{fig:fig1} (a).  Results for different system sizes $L=8,16,24,32,40$ are presented and it is clear that the finite-size effect is negligible. The data clearly demonstrates a linear scaling as in Eq.~\eqref{eq:eq16} and the slope $a_1$ quickly converges to $0.0394 \pm 0.0004$.

\begin{figure}[htp!]
\centering
\includegraphics[width=\columnwidth]{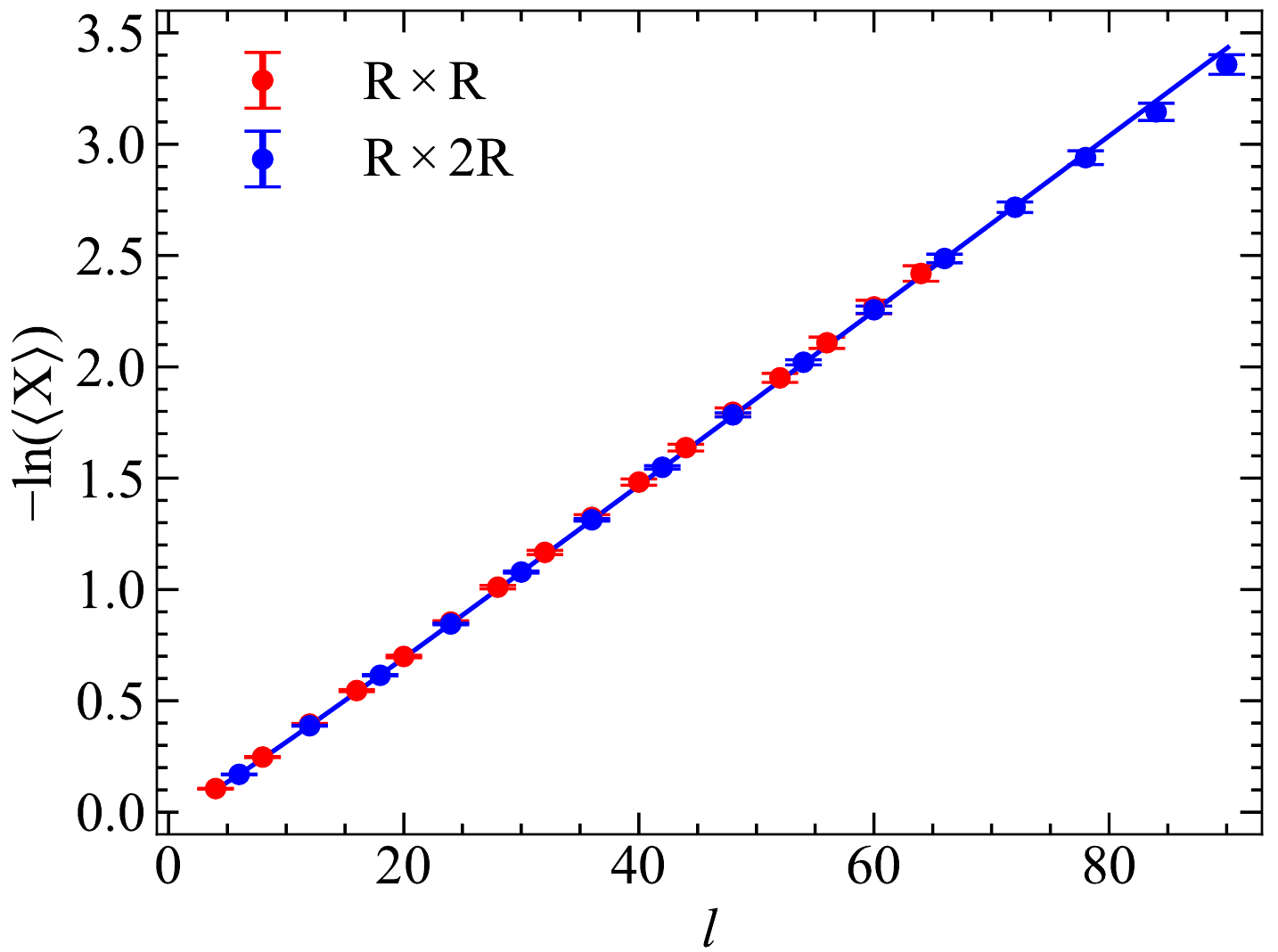}
\caption{$-\ln \langle X_M\rangle$ versus $l$ at the phase transition point for $M$ with the shape $R\times R$ (already shown in Fig.~\ref{fig:fig4}) and $R \times 2R$, for system size $L=32$. The blue line represents the fitted curve of the data for $R\times 2R$ using the relation specified in Eq.~\eqref{eq:eq16}. The fitted result of $R\times 2R$ is $-\ln\langle X \rangle=(0.0397 \pm 0.0002)l-(0.0279 \pm 0.003)\ln(l)-(0.0192 \pm 0.006)$ and for $R\times R$ at $L=32$ the result is $-\ln\langle X \rangle=(0.0399 \pm 0.0003)l-(0.0272 \pm 0.004)\ln(l)-(0.0162 \pm 0.005)$. The coefficients are indistiguishable within errorbars.}
\label{fig:fig5}

\end{figure}

As we have explained, the boundary of $M$ generally contributes to the disorder operator a term proportional to the perimeter. To detect 1-form symmetry breaking, we need to check whether $\langle X\rangle$ depends on the area or not. For this purpose, we consider rectangular regions with different aspect ratios: one with $1:1$ (Fig.~\ref{fig:fig1} (a)) and the other with $1:2$ (Fig.~\ref{fig:fig1} (b)), and present the results of $\langle X \rangle$ at the $h=h_c$ together in Fig.~\ref{fig:fig5}. It can be seen that the two sets of data basically fall on the same curve, indicating that the disorder parameter only depends on the perimeter.

Given the relation between $\langle X_M\rangle$ and the Renyi entropy in the free theory, let us examine possible corner contributions to $\langle X_M\rangle$, which is parameterized in the coefficient $s$ of Eq.~\eqref{eq:eq16}. We fit the data points in Fig.~\ref{fig:fig5} to Eq.~\eqref{eq:eq16}, which yields $s=0.0272 \pm 0.004$, close to the free value. We perform the same fit for data points with aspect ratio $1:2$ and obtain essentially the same results ($s=0.0279\pm 0.003$).  The agreement between the fitting results for regions with different aspect ratios again lends strong support for the perimeter dependence of $\langle X_M\rangle$ even beyond the leading order, and consequently the 1-form symmetry breaking at the $(2+1)$d Ising CFT.

The convergence of the coefficients  $a_1$, $s$ and $a_0$ versus the linear system size $L$ is given in Fig.~\ref{fig:fig7} in Appendix~\ref{Sec:appB3}.

\subsection{Ordered phase $h<h_c$}
\begin{figure}
	\centering
	\includegraphics[width=\columnwidth]{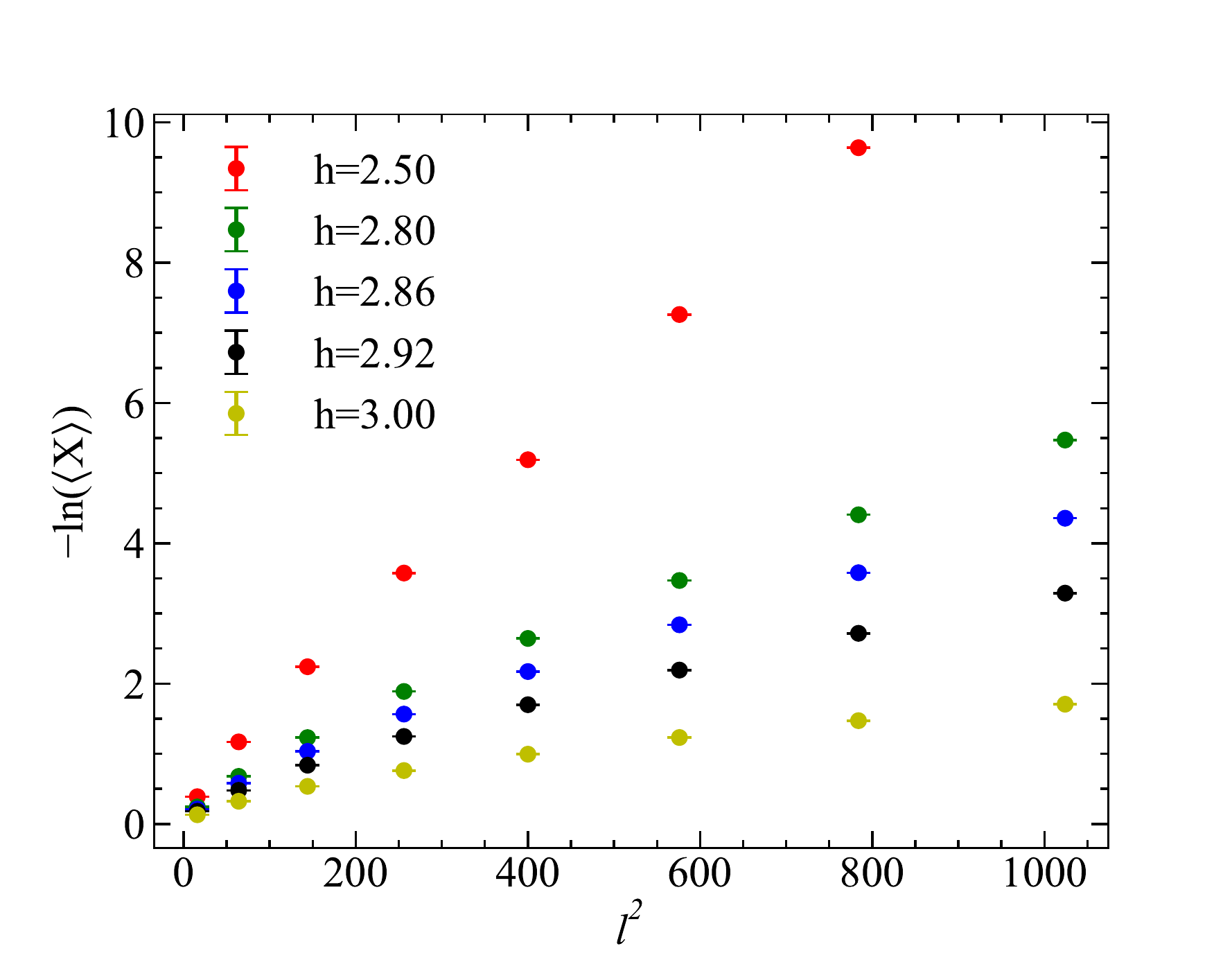}
	\caption{$-\ln \langle X\rangle$ versus $l^2$ when $L=16$ and $\beta=32$. The transverse fields are chosen from inside the ordered phase ($h=2.5$) to near the critical point ($h=3.0$). The area law scaling in the disorder operator clearly manifests, and the slope increases as as one moves deeper in the ferromagnetic phase.}%We use straight line to fit the data of $h=2.80$ and $h=2.80$. The fitted curves are separately 0.0127(3) and 0.0056(2). For $h=2.50$ when $l^2=32^2$ the $\langle X\rangle$ is smaller than the errors so this point is neglected on the figure.}
	\label{fig:fig6}
\end{figure}

For $h<h_c$ where Ising spins order ferromagnetically, our algorithm becomes inefficient because we choose to work in the $\sigma^x$ basis to facilitate the computation of the disorder operator. Nevertheless, simulations indeed find that the disorder parameter decays much more rapidly with the linear size of the region, consistent with the area law in Eqs.~\eqref{eq:eq5} and \eqref{eq:eq9}. $-\ln \langle X_M\rangle$ as a function of $l^2$ is shown in Fig.~\ref{fig:fig6} for different values of $h$ below the critical value. It is clear that as we go deep into the ordered phase, the slope $b$ increases as expected and the data points converge to a straight line for large $l^2$. For $h=3.0$ very close to the critical point, one can observe that for relatively small values of $l^2$ the data points do not scale linearly, which can be attributed to a subleading perimeter dependence. 

\section{Conclusion and Discussion}
\label{sec:v}
As discussed in the beginning of the paper, in recent years, new types of quantum phases and phase transitions that are ``beyond Laudau'' are flourishing, exhibiting topological order, emergent gauge field and fractionalization. Higher-form symmetries and their spontaneous breaking are new conceptual tools introduced to provide an unified framework for both conventional and exotic phases. A quantum phase, gapped or gapless, is fundamentally characterized by its emergent symmetry and the associated anomaly. While the philosophy went back to the Landau classification of phase transitions, the power of this perspective has only begun to unfold recently with the introduction of generalized global symmetries.

Here we re-examine the familiar Ising symmetry-breaking transition, arguably the simplest conformal field theory, from the emergent symmetry perspective. A $D$-dimensional Ising system has a ``hidden'' $\Z_2$ $(D-1)$-form symmetry, whose charges are Ising domain walls. Gapped phases in this system are associated to the spontaneous breaking of the enlarged symmetry (0-form and $(D-1)$-form symmetries). It is then of great interest to determine the symmetry breaking pattern at the critical point, to complete our understanding of the global phase diagram from the emergent symmetry point of view.

In this work we determine the scaling form of the disorder operator in Ising CFTs when $D>1$. The most challenging case is $D=2$ where the transition is described by the interacting Wilson-Fisher fixed point, and we exploit large-scale quantum Monte Carlo simulations. We use the disorder operator of the Ising system to probe the breaking of the dual higher-form symmetry. We find numerically that at the critical point of the 2D quantum Ising model, the one-form disorder operator exhibits sponatenous symmetry breaking as in the disordered phase, whereas in the ordered phase, the one-form symmetry is intact.  

The disorder operator is intimately related to a line defect (also called a twist operator) in a Ising CFT, around which the spin operator sees an anti-periodic boundary condition. In fact, a line defect is nothing but the boundary of a disorder operator. It is believed that in general such a line defect can flow to a conformal one at low energy, which is indeed consistent with a perimeter law scaling for the expectation value of the disorder operator~\footnote{We are grateful for Shu-Heng Shao for discussions on this point.}. Local properties of disorder line defects have been previously investigated in Ref. [\onlinecite{Billo2013}] and [\onlinecite{Gaiotto2013}]. It will be interesting to understand the relation between the local properties with the universal corner contributions to the disorder operator~\cite{Bueno2015}.

Our findings, besides elucidating the physics of quantum Ising systems from a new angle, provides a working example of higher-form symmetry at practical use. 
Similar physical systems can be studied, for example, the disordered operator constructed in this work is readily generalized to the $(2+1)$d XY transition and can be measured with unbiased QMC simulations. Another important direction is to study other higher-form symmetry breaking transitions, such as 1-form symmetry breaking transition in 3D systems. It would also be interesting to investigate the ultility of the disorder operator in the topological Ising paramagnetic phase.
More applications in quantum lattice models are awaiting to be explored, and will certainly lead to new insight for a new framework that unifies our understanding of the exotic quantum phases and transitions going beyond the Landau paradigm and those within. 

{\it Note added.-} We would like to draw the reader’s attention to few closely related recent works by X.-C. Wu, C.-M. Jian and C. Xu~\cite{XCWu2020,XCWu2021} and by some of the present author on scaling of disorder oeprator at $(2+1)$d U(1) quantum criticality~\cite{YCWang2021}. 

\section*{Acknowledgement}
JRZ, ZY and ZYM thank the enjoyable discussions with Yan-cheng Wang and Yang Qi and acknowledge the support from the RGC of Hong Kong SAR of China
(Grant Nos. 17303019 and 17301420), MOST through the
National Key Research and Development Program (Grant
No. 2016YFA0300502) and the Strategic Priority Research
Program of the Chinese Academy of Sciences (Grant No.
XDB33000000). We are grateful for Xiao-Gang Wen, Shu-Heng Shao and William William-Krempa for helpful comments. MC would like to thank Zhen Bi, Wenjie Ji and Chao-Ming Jian for enlightening discussions and acknowledges support from NSF (DMR-1846109). We thank the Computational Initiative at the
Faculty of Science and the Information Technology Services at the University of Hong Kong, and the
Tianhe-1A, Tianhe-2 and Tianhe-3 prototype platforms at the
National Supercomputer Centers in Tianjin and Guangzhou for
their technical support and generous allocation of CPU time.

\appendix

%\onecolumngrid
\section{Quantum Monte Carlo implementation of disorder operator}
\label{sec:app2}

In this appendix, we describe the implementation of SSE-QMC algorithm of the quantum Ising model, in particular the implementation of the disorder operator which involves a change of basis.
\subsection{SSE on $\sigma^{z}$ basis}
The Hamiltonian for the transverse field Ising model is
\begin{equation}
H=-\sum_{\langle \mb{r}\mb{r'}\rangle}\sigma_{\mb{r}}^{z}\sigma_{\mb{r'}}^{z}-h\sum_{\mb{r}}\sigma_{\mb{r}}^{x}.
\end{equation}
Then we can decompose the Hamiltonian into site and bond operators
\begin{equation}
\begin{split}
H_{0, 0}&=I\\
H_{-1,a}&=h(\sigma_{a}^{+}+\sigma_{a}^{-})\\
H_{0,a}&=h\\
H_{1,a}&=(\sigma_{\mb{r}(a)}^{z}\sigma_{\mb{r'}(a)}^{z}+1)\\
\end{split}
\end{equation}
with $H=-\sum_{i=-1}^{1} \sum_{a} H_{i,a}$. Here $H_{0,0}$ denotes the identity operator and $i=-1,0,1$ indicates different types of operator: off-diagonal operator on site, diagonal operator on site and diagonal operator on bond. The subscript $a$ holds two different identities: for bond operators $H_{1,a}$ index $a$ denotes the bond number (e.g. for 2D case $a=1,2,...,N_b=2L^2$); and for site operators $H_{0,a}$ and $H_{-1,a}$ index $a$ denotes the site number (e.g. for 2D case $a=1,2,...,N=L^2$).

Next, the partition function $Z=\mathrm{Tr}\,e^{-\beta H}$ can be expressed as a power series expansion:
\begin{equation}
Z = \sum\limits_\alpha \sum_{S_M} {\beta^n(M-n)! \over M!}
    \left \langle \alpha  \left | \prod_{i=1}^M H_{a_i,p_i}
    \right | \alpha \right \rangle ,
\label{zm}
\end{equation}
where $M$ is the truncation of the expansion series $n$. Taking $\sigma^{z}$ as a complete set of basis for the system, the non-zero matrix elements for site operators and bond operators are
\begin{equation}
\begin{split}
\langle \uparrow|H_{-1,a}|\downarrow\rangle &= \langle \downarrow|H_{-1,a}|\uparrow\rangle=h\\
\langle \uparrow|H_{0,a}|\uparrow\rangle &= \langle \downarrow|H_{0,a}|\downarrow\rangle=h\\
\langle \uparrow\uparrow|H_{1,a}|\uparrow\uparrow\rangle &=\langle \downarrow\downarrow|H_{1,a}|\downarrow\downarrow\rangle=2\\
\end{split}
\end{equation}

\begin{figure}[htp!]
	\centering
	\includegraphics[width=0.9\columnwidth]{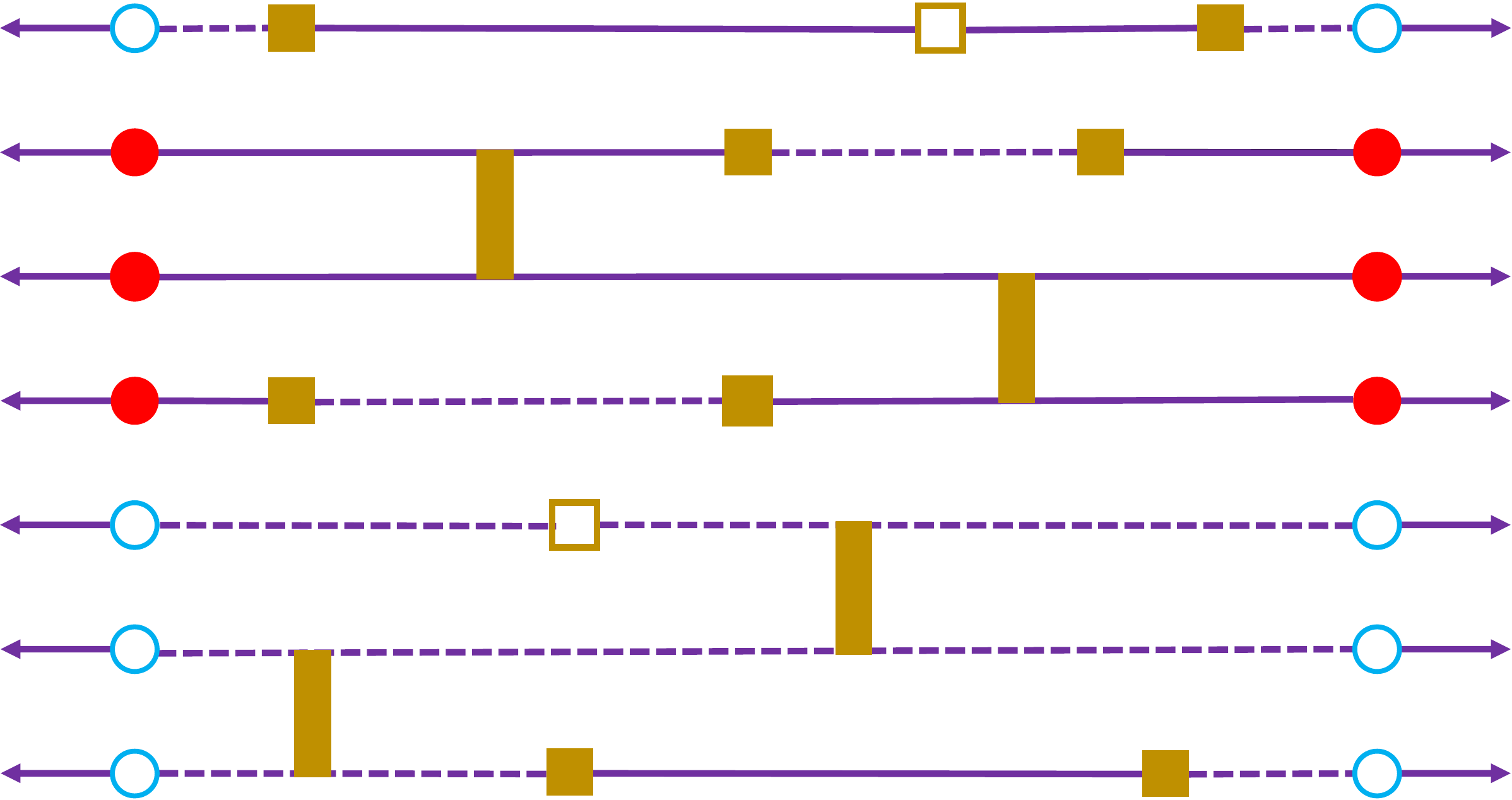}
	\caption{SSE-QMC configuration of quantum Ising model. Golden bars represent Ising bond operators. Filled squares plaquette are off-diagonal site operators, and open plaquettes denote the diagonal site operators. Arrows represent periodic boundary conditions in the imaginary time direction. The red solid circles and the light blue open circles indicate spin up and down. Solid and dashed purple lines illustrate the spin states (spin up or down).}
	\label{fig:fig7}
\end{figure}

The updating scheme~\cite{sandvikTFIM} includes the diagonal update which either inserts or removes a diagonal operator between two states with probabilities regulated by the detailed balanced condition, and the cluster update which flips all the spins on the cluster with the Swendsen-Wang Scheme. The configurations of the updating scheme are shown in Fig.~\ref{fig:fig7}.

We describe the updating scheme in the following steps:
\begin{enumerate}
	\item \textbf{Diagonal update}\\
	We go through the operator strings and either remove or insert a diagonal operator according to the following procedures.
	\begin{enumerate}
		\item For a diagonal operator ($H_{0,a}$ or $H_{1,a}$),
		we removed  it with probability, \begin{equation}
		P=\min\left(\frac{M-n+1}{\beta(hN+2N_{b})},1\right)
		\end{equation}
		where $N$ denotes the number of lattice sites, and $N_b$ denotes the number of bonds.
		\item For a null operator ($H_{0,0}$) , we substitute it with a diagonal operator $H_{1,a}$ or  $H_{0,a}$  by the procedures below.
		\begin{enumerate}
			\item Firstly we make the decision of which kind of diagonal operators to insert. We choose the type of $H_{1,a}$ with probability \begin{equation}
			P(h)=\frac{2N_{b}}{hN+2N_{b}}
			\end{equation} or the type $H_{0,a}$ with probability $1-P(h)$.
			\item After the decision is made, accept the insertion of an operator with probability
			\begin{equation}
			P=\min\left(\frac{\beta(hN+2N_{b})}{M-n},1\right),
			\end{equation}
			and after that we choose an random and appropriate site or bond to insert the operator. If the chosen bond to insert a bond operator has an anti-parallel configurations, then the insertion  of a bond operator at this place is prohibited.
		\end{enumerate}
	\item For an off-diagonal operator, we ignore it and go to the next operator in the operator strings.
	\end{enumerate}
	\item \textbf{Cluster update}
	\begin{enumerate}
		\item We generally follow two rules to construct the clusters: (1) clusters are terminated
		on site-operators $H_{-1,a}$ or	$H_{0,a}$; and (2) the four legs of a bond operator
		$H_{1,a}$ belong to one cluster.
		Carry out this procedure until all the clusters are bulit, and a configuration of clusters are shown in Fig.~\ref{fig:fig7}.
		\item Clusters identified from the above rules are then flipped with probability $1/2$ (which is the Swendsen Wang cluster updating scheme).
	\end{enumerate}
	
\end{enumerate}

Since the disorder operator is a product of $\sigma^{x}$, i.e., $\langle X_{M} \rangle = \langle \prod_{\mb{r}\in M}\sigma_\mb{r}^x \rangle$, it is a measurement of an off-diagonal operator in the $\{\sigma^{z}\}$ basis. In the $\sigma^{z}$ basis, the off-diagonal operator can be measured if the operator is a product of operators in the Hamiltonian. It is proved in Ref. \onlinecite{OFS2002} that
\begin{equation}
\left\langle\prod_{i=1}^{m} \hat{H}_{k_{i}}\right\rangle=\frac{1}{(-\beta)^{m}}\left\langle\frac{(n-1) !}{(n-m) !} N\left(k_{1}, \ldots, k_{m}\right)\right\rangle_{W}
\end{equation}
where $N(k_1,\dots,k_m)$ denotes the number of ordered sub-sequences $k_1,...,k_m$ in $S_{n}$. However this measurement becomes practically impossible when the length of the products becomes sufficiently large, because $\frac{1}{(-\beta)^{m}}\frac{(n-1) !}{(n-m) !}$ would grow to a very large value as $m$ increases, thus  $N(k_1,...,k_m)$ would be too small to measure within the limited computing power. So the measurement of $\langle X \rangle$ in the $\{\sigma^{z}\}$ basis seems hopeless. To solev this problem, we need to change the basis to make $\sigma^{x}$ diagonal.

\subsection{SSE on $\sigma^{x}$ basis}
Since we need to measure the disorder operator which is defined as the non-local product of off-diagonal operators, it is extremely hard to measure it in the traditional $\sigma^{z}$ basis, we then turn to $\sigma^{x}$ basis as the complete set of basis of the system, and we can use directed loop algorithms~\cite{OFS2002,OFS2003} to simulate this model.

For convenience, we now write the $\sigma^{x(z)}$ above as $\sigma^{z(x)}$ in following, the Hamiltonian can be rewritten as,
\begin{equation}
H=-\sum_{\langle \mb{r}\mb{r'}\rangle}\sigma_{\mb{r}}^{x}\sigma_{\mb{r'}}^{x}-h\sum_{\mb{r}}\sigma_{\mb{r}}^{z}+N_b\Delta.
\end{equation}

Here $\langle \mb{r}\mb{r'}\rangle$ refers to the nearest neighbors. $N_b$ is the number of bonds. $N_b\Delta$ is a constant added to the Hamiltonian to ensure that the matrix elements defined in Eq.~\eqref{off-matrix} are positive definite. Rewriting $S^x$ with $S^++S^-$, we can decompose the Hamiltonian as
\begin{equation}
H=- \sum_{b=1}^{N_{b}} H_{b},
\end{equation}
with
\begin{equation}
H_{b}=-H_{1, b}-H_{2, b}+H_{3,b}.
\end{equation}
Here $b$ refers to1 the bond number, and $H_{1,b}, H_{2,b}, H_{3,b}$ are defined as follows:
\begin{equation}
\begin{split}
H_{1,b}&=\sigma_{\mb{r}(b)}^{+}\sigma_{\mb{r'}(b)}^{+}+\sigma_{\mb{r}(b)}^{-}\sigma_{\mb{r'}(b)}^{-}\\
H_{2,b}&=\sigma_{\mb{r}(b)}^{+}\sigma_{\mb{r'}(b)}^{-}+\sigma_{\mb{r}(b)}^{-}\sigma_{\mb{r'}(b)}^{+}\\
H_{3,b}&=\Delta-ah(\sigma_{\mb{r}(b)}^{z}+\sigma_{\mb{r'}(b)}^{z}).
\end{split}
\end{equation}
Note that $a=\frac{N}{2N_{b}}$ and $N$ is the number of lattice sites. For the 1D case $a=\frac{1}{2}$, and for the 2D case $a=\frac{1}{4}$.
The non-zero matrix elements for the diagonal operators are
\begin{equation}
\label{diagonal-marix}
\begin{split}
\langle\uparrow\uparrow|H_{b}|\uparrow\uparrow\rangle&=\Delta-2ah\\
\langle\downarrow\downarrow|H_{b}|\downarrow\downarrow\rangle&=\Delta+2ah\\
\langle\uparrow\downarrow|H_{b}|\uparrow\downarrow\rangle&=\langle\downarrow\uparrow|H_{b}|\downarrow\uparrow\rangle=\Delta,	\\
\end{split}
\end{equation}
In the simulation we set $\Delta=2ah+1$ to make sure $H_{b}$ is positive definite.
The off-diagonal matrix elements are
\begin{equation}
\label{off-matrix}
\langle\uparrow\uparrow|H_{b}|\downarrow\downarrow\rangle=\langle\downarrow\downarrow|H_{b}|\uparrow\uparrow\rangle=\langle\uparrow\downarrow|H_{b}|\downarrow\uparrow\rangle=\langle\downarrow\uparrow|H_{b}|\uparrow\downarrow\rangle=1
\end{equation}

Then the updating scheme becomes:
\begin{enumerate}
	\item \textbf{Diagonal update}\\
	The purpose of diagonal update which either inserts or removes a diagonal between two basis states is to change the expansion order $n$ by $\pm 1$. The corresponding acceptance probability is
	\begin{equation}
		\begin{split}
			P\left( \text{insert} \right)&=\frac{N_{b} \beta\left\langle\alpha(p)\left|H_{1, b}\right| \alpha(p)\right\rangle}{M-n} \\
			P\left( \text{remove} \right)&=\frac{M-n+1}{N_{b} \beta\left\langle\alpha(p)\left|H_{1, b}\right| \alpha(p)\right\rangle}
		\end{split}
\end{equation}

	\item \textbf{Directed loop update}\\
	We can construct the loop as following. Firstly, select randomly one of vertex legs as an initial entrance leg. Exit vertex leg is chosen with the probability as Eq.~\eqref{PPP}, and both the entrance and exit spins are flipped.
	The probability of exit leg is defined with matrix elements obtained by flipping spins in a vertex. The elements are defined as
	\begin{equation} W\left(\begin{array}{l} g_{3}, g_{4} \\ g_{1}, g_{2} \end{array}\right) =\left\langle g_{3} S_{i}^{z}, g_{4} S_{j}^{z}\left|H_{b}\right| g_{1} S_{i}^{z}, g_{2} S_{j}^{z}\right\rangle 
	\end{equation}
	
	where $g_{i}=-1$ if the spin on leg $i$ is flipped and $g_{i}=+1$ if
	it is not flipped. For example the probability of exiting at leg 3
	if the entrance is at leg 1 is given by
	\begin{equation}\label{PPP}P_{3,1}=\frac{W\left(_{-+}^{-+}\right)}{W\left(_{++}^{++}\right)+W\left(_{--}^{++}\right)+W\left(_{-+}^{-+}\right)+W\left(_{-+}^{+-}\right)}\end{equation}
Then let it visit next vertex. The loop goes on in this way one vertex by one until it closes. Also we use the Swendsen Wang scheme to flip the clusters after all clusters are identified. \\
\end{enumerate}

\begin{figure}
\centering
\includegraphics[width=\columnwidth]{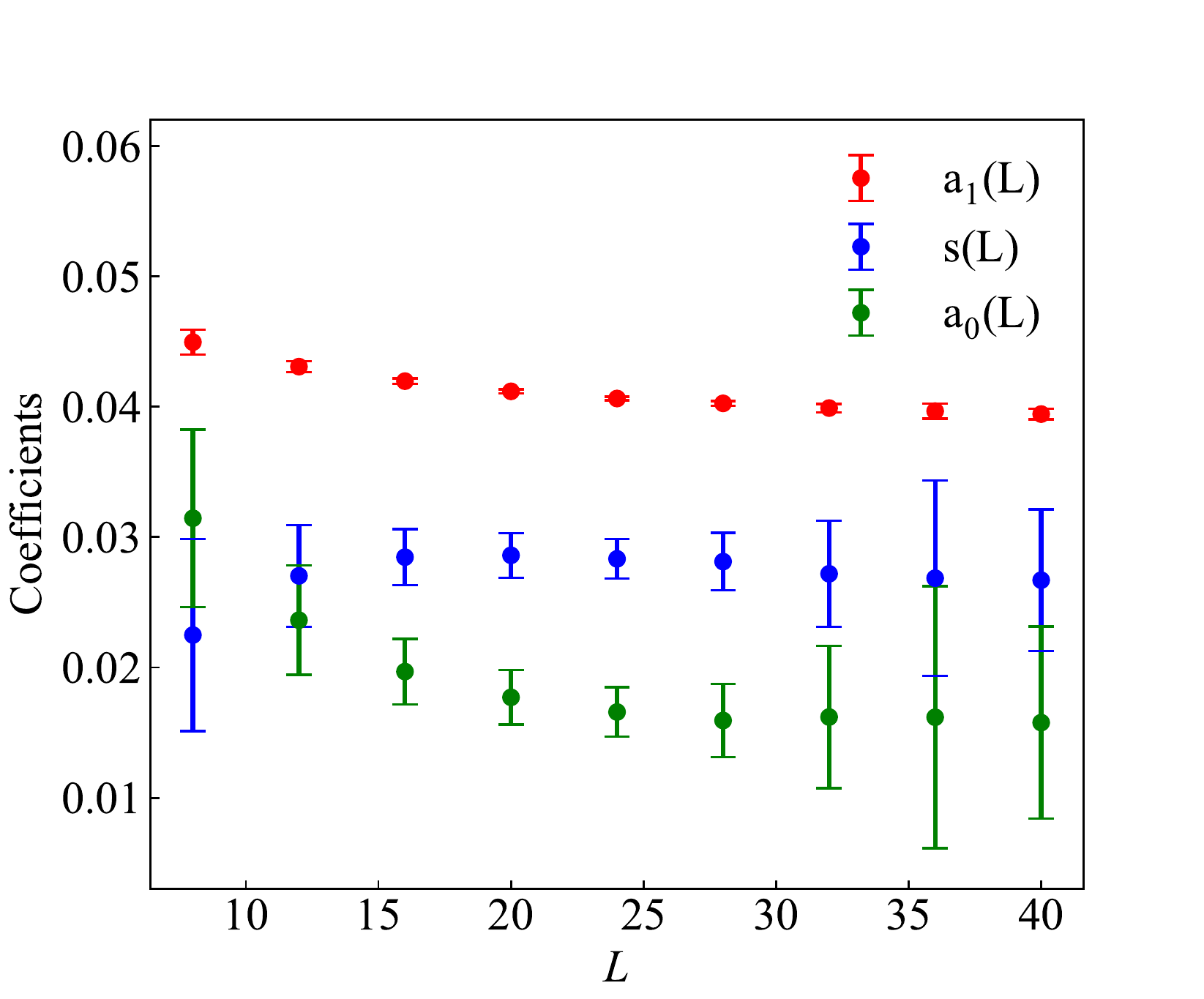}
\caption{Finite-size convergence of the coefficients of the disorder operator at the $(2+1)$d Ising critical point, for the case of $M=R\times R$.}
\label{fig:fig8}
\end{figure}

\subsection{Curve fitting}
\label{Sec:appB3}
Lastly, we show the details of fitting results of Fig.~\ref{fig:fig4} and ~\ref{fig:fig5}. We fit the disorder operator according to Eq.~\eqref{eq:eq16} and obtained the coefficents $a_1$, $s$ and $a_0$ for different system sizes. Fig.~\ref{fig:fig8} demonstrate the convergence of the fitting results as the system size increases.

\twocolumngrid

\bibliography{cft}
\end{document}